\documentclass[fleqn,usenatbib]{mnras}

\usepackage{newtxtext,newtxmath}

\usepackage[T1]{fontenc}

\DeclareRobustCommand{\VAN}[3]{#2}
\let\VANthebibliography\thebibliography
\def\thebibliography{\DeclareRobustCommand{\VAN}[3]{##3}\VANthebibliography}

\usepackage{graphicx}	
\usepackage{amsmath}	
\usepackage{amssymb}	

\title[IC1459 Group UDGs]{Ultra Diffuse Galaxies in the IC1459 Group from the VEGAS Survey}

\author[D. A. Forbes et al.]{
Duncan A. Forbes,$^{1}$\thanks{E-mail: dforbes@swin.edu.au (DAF)}
Bililign T. Dullo,$^{2}$
Jonah Gannon$^1$, Warrick J. Couch$^1$, Enrichetta Iodice$^3$
\newauthor
 Marilena Spavone$^3$, Michele Cantiello$^4$ and Pietro Schipani$^3$
\\
$^{1}$Centre for Astrophysics \& Supercomputing, Swinburne University, Hawthorn VIC 3122, Australia\\
$^{2}$Departamento de Fısica de la Tierra y Astrofısica, Instituto de Fısica de Part´ıculas y del Cosmos IPARCOS, Universidad Complutense
de Madrid, E-28040, Spain\\
$^{3}$INAF-Astronomical Observatory of Capodimonte
Salita Moiariello 16, 80131, Naples, Italy\\
$^4$INAF Osservatorio Astronomico d’Abruzzo, via Maggini, snc, I-64100, Italy
}

\date{Accepted XXX. Received YYY; in original form ZZZ}

\pubyear{2015}

\begin{document}
\label{firstpage}
\pagerange{\pageref{firstpage}--\pageref{lastpage}}
\maketitle

\begin{abstract}
Using deep g,r,i imaging from the VEGAS survey, we have searched for ultra diffuse galaxies (UDGs) in the IC 1459 group. Assuming they are group members, we identify 9 galaxies with physical sizes and surface brightnesses that match the UDG criteria within our measurement uncertainties. They have mean colours of g--i = 0.6 and stellar masses of  $\sim$10$^8$ M$_{\odot}$. Several galaxies appear to have associated systems of compact objects, e.g. globular clusters. Two UDGs contain a central bright nucleus, with a third UDG revealing a remarkable double nucleus. This appears to be the first reported detection of a double nucleus in a UDG -- its origin is currently unclear.

\end{abstract}




\section{Introduction}

Ultra diffuse galaxies (UDGs) are a class of low surface brightness galaxy with effective radii R$_e$ $>$ 1.5 kpc and central surface brightnesses in the g band fainter than 24 mag. per sq. arcsec. Since their identification by van Dokkum et al. (2015), they have been the subject of many galaxy formation simulations, and recent examples include Liao et al. (2019), Sales et al. (2019), Tremmel et al. (2019) and Martin et al. (2019). 
Although these simulations can explain many properties of UDGs as dwarf galaxies that have been 'puffed-up' to large size, they do not predict the massive halos (van Dokkum et al. 2019) nor the rich globular cluster systems (Forbes et al. 2020) for some UDGs.  
In many of these simulations, UDGs 
are pre-processed in groups in a low density environment and later fall into a cluster.

While the abundance of UDGs is indeed greater in higher density environments (i.e. the number of UDGs scales in a near-linear fashion with halo mass; Janssens et al. 2019), they have also been found in low density environments.  One recent example is five UDG candidates identified by Muller et al. (2018) in the nearby Leo group (D $\sim$ 10 Mpc) region of the sky. To date, none of these five have reported distances. One nearby group (D $\sim$ 18 Mpc) with distance-confirmed UDGs is the NGC 1052 group, i.e. NGC1052--DF2 and NGC1052--DF4 (Blakeslee \& Cantiello 2018; Danieli et al. 2019).

The first step to identifying UDGs in low density environments is deep imaging to low surface brightness levels. The VEGAS survey is ideally suited to this task. Using OmegaCAM on the VST (Capaccioli \& Schipani 2011), VEGAS is imaging nearby early-type galaxies and their surroundings over a range of environments. It reaches down to surface brightness levels of $\sim$29 mag per sq. arcsec in the g band (Spavone et al. 2017; Venhola et al. 2017). 
Published UDG detections from the VEGAS survey have included 9 candidates in the Fornax cluster (Venhola et al. 2017) and one in the NGC 5846 group (Forbes et al. 2019).

Here we report on a search for UDG candidates in the deep VEGAS imaging of the IC 1459 group.
The group has a systemic velocity of 1709 km/s and we assume the same distance as Iodice et al. (2019) of 28.7 Mpc. The group contains a hot  intra group medium as revealed by diffuse X-ray emission (Osmond \& Ponman 2004). Although there are indications of interactions within the group from extended HI emission, several galaxies look undisturbed in terms of their HI content (Kilborn et al. 2009; Serra et al. 2015). 
Based on the measurements of 
Brough et al. (2006), Iodice et al. (2019) calculated a virial radius for the group of 
R$_{200}$ = 0.21 Mpc and a group halo mass of M$_{200}$ = 3.7 $\times$ 10$^{13}$ M$_{\odot}$.
The group 
is dominated by the giant elliptical IC 1459 which exhibits  shells, plumes and dust indicative of a recent merger event (Forbes et al. 1995). 

\section{Analysis}

\subsection{Finding Ultra Diffuse Galaxies}

     \begin{figure*}
   \centering
   \includegraphics[width=16cm,angle=0]{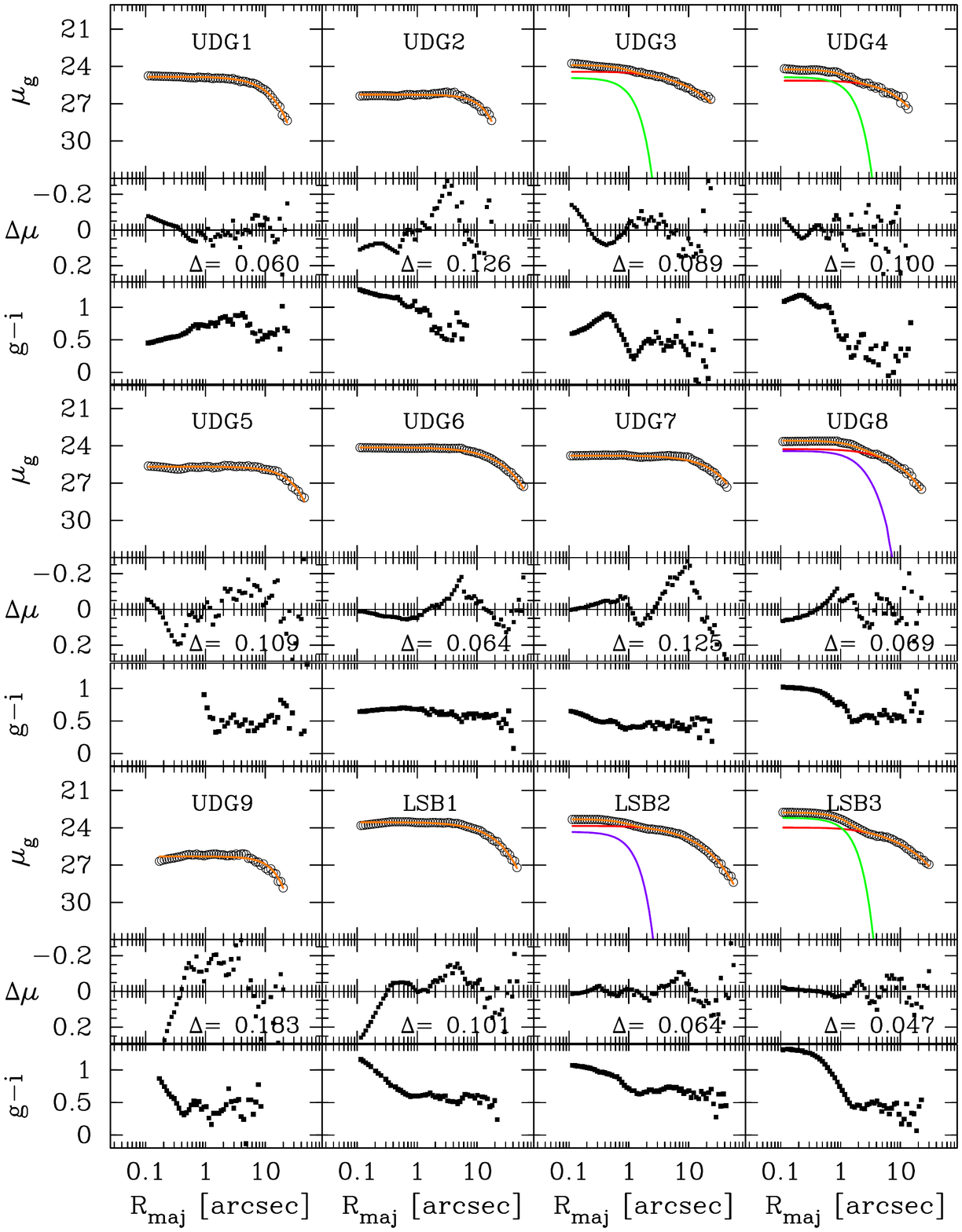}
      \caption{1D profile fits to IC1459 UDG candidates. Panels show the g band surface brightness (open circles) of each UDG or LSB galaxy measured along  their major axis. The central nuclear components are fit by a Gaussian (green) or S\'ersic profile (purple). The main S\'ersic component (red) and the combined fit (orange) are also shown. Panels include residuals to the combined fit and the g--i colour profile.  
                  }
         \label{SB}
   \end{figure*}

The VEGAS imaging and data reduction for the IC1459 group, used here, are described in Iodice et al. (2019). The final mosaic images in the g,r,i bands are 1$\times$2 sq. degree (with 0.21 arcsec pixels). The seeing conditions were 1.73, 0.89, 0.94 arcsec 
in the g,r,i bands, respectively. 
The IC1459 group reveals a wealth of low surface brightness features in the VEGAS imaging (see figure 2 of Iodice et al. 2019). These features include shells and plumes and extended halo light around the galaxies, but also light from the intra group medium. This makes automatic detection of UDGs by an algorithm problematic. Here we have followed the same method as applied successfully to VEGAS imaging of the NGC5846 group (Forbes et al. 2019), of finding UDGs by visual inspection of the images using a high resolution screen. 

We display the mosaic image on a high resolution 4K $\times$ 4K device and examine in subsections so that individual pixels are resolved. We carefully examine each image searching for low surface brightness, large-sized objects with diameters of tens of pixels (corresponding to a few kpc at the distance of the IC1459 group). 
This process is successful at finding 12 low surface brightness galaxies which we refine below to confirm their UDG status by measuring their sizes and central surface brightnesses.

\subsection{Measuring Ultra Diffuse Galaxies} 


Before modelling the UDG candidates, 
we measure the median of the sky values in several 50 $\times$ 50 pixel boxes in a blank region of the image
and subtract it. 
Masks are constructed around each UDG candidate to exclude any compact sources, background galaxies, bright foreground stars and extended stellar haloes.
These mask images are used when we fit 
elliptical isophotes, using 
the IRAF task {\sc ellipse}, 
to the sky-subtracted g,r,i band images of each galaxy. We create a 2D model of each galaxy allowing the position angle and ellipticity to vary. 

From our 2D galaxy model we create a 1D surface brightness profile along the major axis. We fit this profile with a S\'ersic model convolved with a Gaussian point spread function following the method of
Dullo et al. (2019). 
The point spread functions are determined from several stars in each image. 
Of the 12 candidate UDGs, seven are well described by a single S\'ersic model. However, five galaxies contain additional central light above that of a single S\'ersic. 
For these five, we include in the fit an additional S\'ersic or Gaussian model.

Fig.~\ref{SB} shows our final fits to the g band light profiles of each galaxy together with the corresponding rms residuals and g--i colour profiles.  The figure shows that the light profiles of all galaxies are well fit with  
our PSF-convolved  S\'ersic (plus Gaussian/S\'ersic) model, as shown by the small residuals.
To estimate uncertainties on the fit parameters, we create 200 Monte Carlo realizations of the galaxy light profile, accounting for 
 errors from incorrect sky subtraction and for possible contamination due to  bright foreground stars, background galaxies and extended stellar halos. Typical 1$\sigma$ uncertainties are $\pm$0.3 kpc in size, $\pm$0.1 mag per sq. arcsec in surface brightness and total magnitude, and $\pm$0.1 in S\'ersic n value. The final best fit parameters for the 12 galaxies are given in Table 1.

\section{Results and Discussion} 

\begin{figure}
   \centering
   \includegraphics[width=7cm,angle=-90]{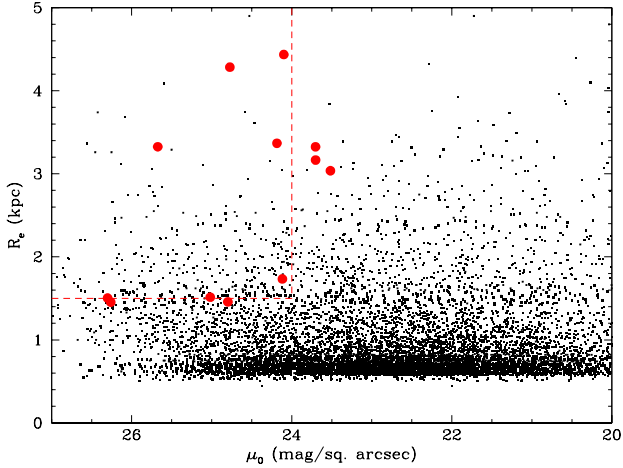}
      \caption{Size vs central surface brightness in the g band. The UDG definition criteria are shown by the dashed lines. IC1459 galaxies are shown by red filled circles. The typical uncertainty is $\pm$ 0.3 kpc in size and $\pm$ 0.1 in surface brightness. We classify 9 galaxies as UDGs and 3 as (slightly brighter) low surface brightness  galaxies. Coma cluster galaxies (from Alabi et al. 2020, in prep.) are shown as small black squares for comparison.}
         \label{coma}
   \end{figure}

From visual inspection of deep imaging of the IC1459 group we identify several large, low surface brightness galaxies. Modelling these objects with a Sersic profile (and an additional  central component in the case of a nucleus) we measure their surface brightnesses and physical sizes (assuming that they lie at the IC1459 group distance) in each of our 3 bands. In Fig.~\ref{coma} we compare their g band properties with galaxies in the Coma cluster, assuming a conversion of g--r = 0.5 (Alabi et al. 2020, in prep.). Assuming a distance of 28.7 Mpc, we find that 7 galaxies are consistent with being a UDG, i.e. R$_e$ $\ge$ 1.5 kpc and $\mu_0$ $>$ 24 mag. per sq. arcsec in the g band (van Dokkum et al. 2015). We emphasise that we use a central surface brightness that excludes the contribution from a central component (if present). 
If we take into account typical uncertainties of $\pm$ 0.3 kpc, or 
assume that they lie slightly beyond 28.7 Mpc, then two more galaxies would be classified as UDGs (these two UDGs have measured g band sizes of $>$ 1.4 kpc). Hereafter, we refer to all 9 galaxies as UDGs. 
Our measurements in  the r and i bands generally agree with those in the g band to within $\pm$ 0.3 kpc. The remaining three galaxies do not meet the UDG criteria as they are slightly brighter in terms of their central surface brightnesses. Hereafter, we refer to these as low surface brightness (LSB) galaxies. 

  Given that a full search of the IC1459 group is hampered by low surface brightness features associated with the intra group medium and galaxy halos, our finding of 9 UDGs could be a lower limit. We note that 
for a group halo mass of M$_{200}$ = 3.7 $\times$ 10$^{13}$ M$_{\odot}$ we expect 5--10 UDGs in the IC1459 group (Janssens et al. 2019). A visual examination the location of the UDGs within the IC1459 group suggests no clear connection with the larger galaxies or the extended HI emission.

The g--i colour profiles of the UDGs (Fig.~\ref{SB}) typically become bluer with radius. This suggests a negative metallicity gradient is present. The UDGs have a mean colour of $<$g--i$>$ $\sim$ 0.6. 
This can be compared to the simulations of 
UDGs 
by Liao et al. (2019). They found a bimodal colour distribution with peaks around g--i = 0.5 and 0.9; with the blue UDGs found in lower density environments than the red UDGs. Thus the IC1459 UDGs have colours more similar to the model UDGs in low density environments. 
Their model UDGs have S\'ersic indices with a wide range of values from around 0.25 to 1.25, with a median value of n = 0.83. We measure mean S\'ersic index of $<$n$>$ $\sim$ 0.85 for the IC1459 UDGs. We also find a mean ellipticity of $\sim$0.25. 

In Table 2 we list the coordinates of each galaxy, along with its total apparent magnitude in the g and i bands. 
Total magnitudes in the AB system were calculated by integrating the best-fitting S\'ersic model parameters. They are not corrected for Galactic extinction which is small, e.g. A$_i$ = 0.025. 
Assuming M$_g$ = +5.11 
in the AB system for the Sun (Willmer et al. 2018) and M/L = 1, the total stellar masses for the UDGs range from 10$^7$ to 4 $\times$ 10$^8$ M$_{\odot}$ (which is similar to that of Coma cluster UDGs; Forbes et al. 2020). 

\begin{figure}
   \centering
   \includegraphics[width=8cm,angle=0]{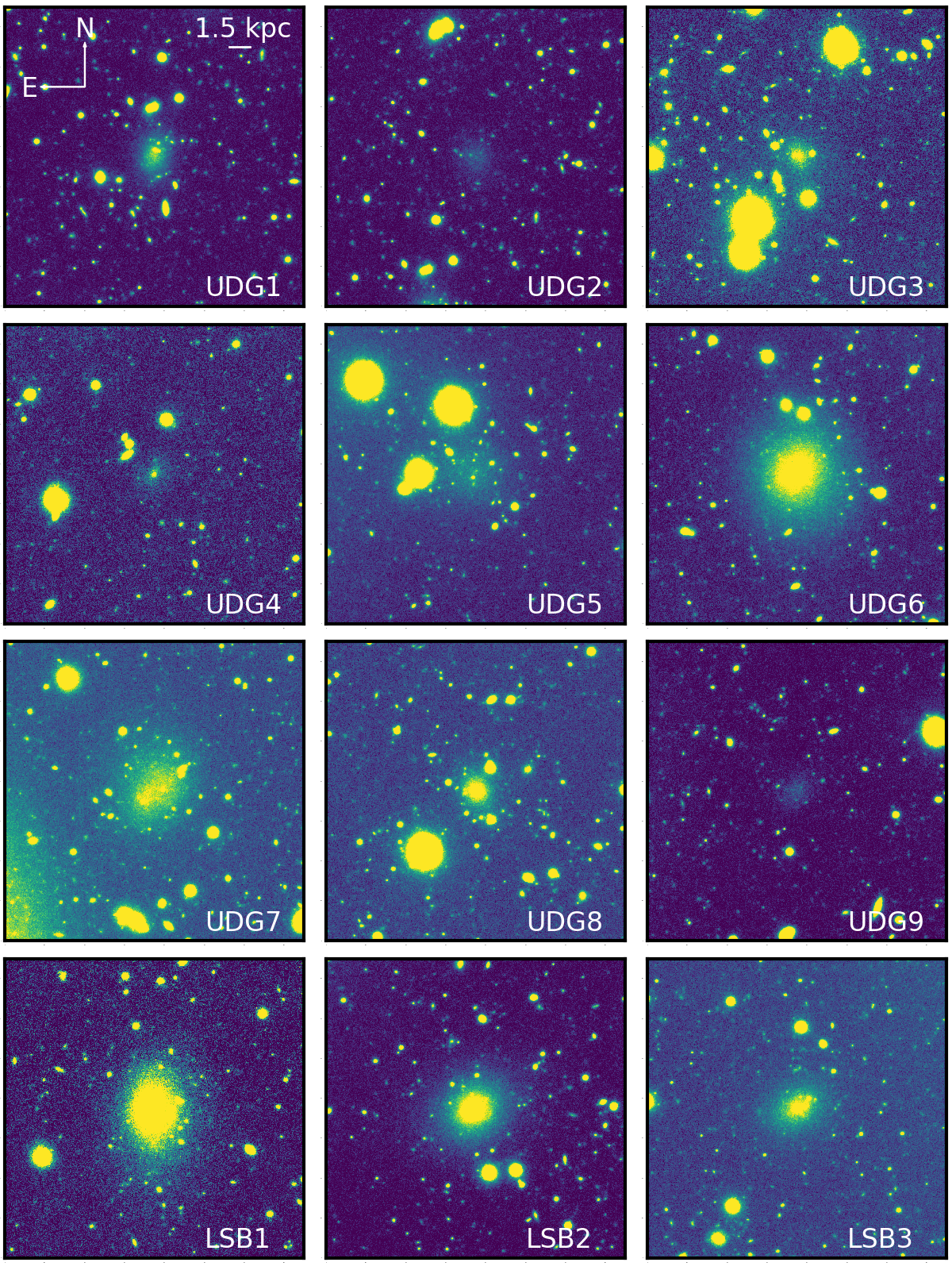}
      \caption{VEGAS images of IC1459 UDG candidates. Each image shows a 158 $\times$ 158 arcsec stacked g+r+i image cutout from the larger VEGAS mosaic image, centred on the UDG. A 1.5 kpc scale is shown assuming a distance of 28.7 Mpc.              }
         \label{image}
   \end{figure}

In Fig.~\ref{image} we show cutout images of all 12 galaxies. The galaxies do not show obvious signs of disturbance but several do reveal the presence of a bright nucleus, which we have fit with an additional central Gaussian or S\'ersic component (along with the standard S\'ersic component for the main galaxy). In particular, we find UDGs 3,4 and 8 to have central nuclei. The presence of central nuclei in UDGs is not uncommon (e.g. Janssens et al. 2019; Alabi et al. 2020, in prep.). 
From the 1D profiles (Fig.~\ref{SB}), the nuclei are redder than the surrounding main galaxy suggesting either metal-rich and/or old nuclei. 
We also find nuclei in two LSB galaxies, i.e LSB2 and LSB3. Thus 3/9 UDGs and 2/3 LSBs show evidence for nuclei. 

\begin{figure}
   \centering
   \includegraphics[width=8cm,angle=0]{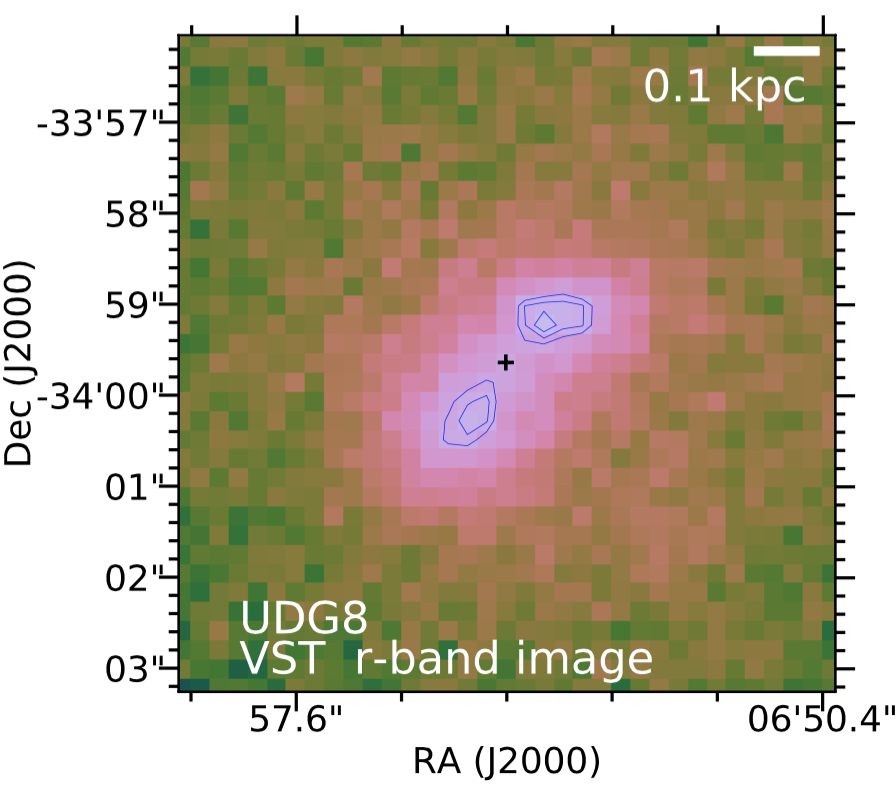}
      \caption{Zoomed r band image of UDG8 with contours shown in blue. The galaxy reveals a double nucleus, with the photometric centre of the galaxy indicated by a small black cross. A 0.1 kpc scale is shown assuming a distance of 28.7 Mpc.              }
         \label{double}
   \end{figure}

Closer examination of IC1459$\_$UDG8 (see Fig.~\ref{double}) reveals it to have a double nucleus, giving it a peanut shape with a central deficit of light (although one nucleus is slightly brighter than the other). The outer isophotes of the galaxy are centred between the two nuclei, suggesting that neither nucleus is the dynamical centre of the galaxy.  This deficiency of light between the nuclei is not associated with redder colours and so unlikely to be the result of a dust lane. It is possible that both nuclei are foreground and/or background objects but their positioning and alignment make this extremely unlikely. Their red colours suggest that the nuclei are made up of old stars rather than hosting AGN. They both have absolute magnitudes of M$_g$ $\sim$ --10 corresponding to a stellar mass of $\sim$10$^6$ M$_{\odot}$ each. 
As far as we are aware this is the first time a double nucleus has been found in a UDG.

Double nuclei in large galaxies are relatively common, with the true nucleus usually lying at the photometric centre of the galaxy. 
For the giant elliptical galaxy NGC 5419, Mazzalay et al. (2016) suggest that the two nuclei both host massive black holes -- one at the photomeric centre of NGC 5419 itself and the other in a disrupted galaxy currently being accreted. 
Closer to home, M31 hosts a double nucleus (Lauer et al. 1998) which Bender et al. (2005) have suggested are central disks viewed at a particular angle. Gas flows along a bar, via secular evolution, may also contribute to peanut-like structures
(see Saglia et al. 2018 on the peanut bulge in M31). 

For UDG8 it would be odd that accreted another galaxy with a nucleus of similar brightness to its own, 
and otherwise show little sign of disturbance. The masses of the two nuclei are however similar to globular clusters -- perhaps we are witnessing the final stages of two globular clusters sinking and merging to the centre of the UDG. Projection effects arising from a torus/disk remain a possibility, but must be a rare occurrence in UDGs. As there is no evidence for a bar structure in UDG8 (or other UDGs we are aware of), it is unlikely that the peanut structure is driven by secular evolution. The origin of the double nucleus in UDG8 thus remains unclear.


Globular clusters (GCs) at the distance of the IC1459 group are not resolved in $\sim$1 arcsec seeing (GCs have typical sizes of 2--3 pc). Nevertheless we have visually inspected the images of Fig.~\ref{image} to identify compact sources around each galaxy. Comparison with similar sized random fields in the group outskirts gives a qualitative measure of whether the galaxy has many, or few, associated compact sources, i.e. GC rich or GC poor. We find UDGs 5, 6 and 7, (along with LSBs 1 and 2) to be potentially GC rich.  Interestingly, these three UDGs also have large physical sizes (see Table 1).
Imaging with the Hubble Space Telescope, which can partially resolve GCs at the distance of the IC1459 group, is perhaps the best approach to more quantitatively measure their GC systems.

   \begin{table}
      \caption{S\'ersic fit parameters}
         \label{KapSou}
            \begin{tabular}{l c c c c c c}
            \hline
            \noalign{\smallskip}
            ID      & band & $\mu_e$ & $< \mu >$$_e$ & R$_e$ & n & $\mu_0$\\
            \noalign{\smallskip}
            \hline
            \noalign{\smallskip}
            
  UDG1 &g  &26.049  &25.488   &1.46   &0.73  &24.798 \\
   &r  &25.691  &25.110   &1.53   &0.77 &24.365 \\ 
   &i  &25.356  &24.812   &1.45   &0.70  &24.169  \\
  UDG2 &g  &27.044  &26.625   &1.45   &0.51  &26.261  \\
   &r  &26.544  &26.113   &1.62   &0.53  &25.724  \\
   &i  &26.487  &25.917   &1.55   &0.75  &25.205  \\
  UDG3 &g  &26.950  &26.079   &3.37   &1.43  &24.185  \\
   &r  &26.817  &25.853   &3.96   &1.73  &23.413  \\
   &i  &26.142  &25.328   &2.24   &1.27  &23.724  \\
  UDG4 &g  &26.822  &26.127   &1.51   &0.99  &25.018  \\
   &r  &26.583  &25.703   &1.28   &1.46  &23.766  \\
   &i  &27.198  &26.243   &2.04   &1.70 &23.859  \\
  UDG5 &g  &26.716  &26.215   &3.33   &0.63  &25.674  \\
   &r  &25.994  &25.696   &2.31   &0.35  &25.551  \\
   &i  &25.954  &25.526   &2.87   &0.52  &25.144  \\
  UDG6 &g  &25.848  &25.164   &4.44   &0.96  &24.099  \\
   &r  &25.360  &24.669   &4.41   &0.98  &23.576  \\
   &i  &25.364  &24.639   &4.29   &1.06  &23.417  \\
  UDG7 &g  &26.300  &25.668   &4.28   &0.86  &24.773  \\
   &r  &25.788  &25.215   &3.73   &0.75  &24.495  \\
   &i  &25.492  &24.939   &3.13   &0.72  &24.271  \\
     UDG8 &g  &26.115  &25.380   &1.73   &1.08  &24.117  \\
   &r  &25.722  &24.926   &1.72   &1.23  &23.404  \\
   &i  &25.600  &24.830   &1.79   &1.16  &23.427  \\
  UDG9 &g  &27.042  &26.637   &1.50   &0.49  &26.301  \\
   &r  &26.434  &26.176   &0.98   &0.24  &26.193  \\
   &i  &26.784  &26.280   &1.29   &0.64 &25.733  \\
   \hline
     LSB1 &g  &25.123  &24.472   &3.04   &0.90  &23.517  \\
   &r  &24.598  &23.976   &2.81   &0.84  &23.113  \\
   &i  &24.595  &23.914   &3.17   &0.96  &22.858  \\
  LSB2 &g  &26.043  &25.243   &3.17   &1.24 &23.702  \\
   &r  &25.577  &24.755   &3.08   &1.29  &23.111  \\
   &i  &25.273  &24.480   &2.75   &1.22  &22.978  \\
  LSB3 &g  &26.595  &25.704   &3.33   &1.49  &23.703  \\
   &r  &25.949  &25.127   &2.27   &1.29  &23.485  \\
   &i  &25.998  &25.102   &2.59   &1.51  &23.072  \\
            \noalign{\smallskip}
            \hline
\end{tabular}
Notes: 
Typical 1$\sigma$ uncertainties are $\pm$0.3 kpc in effective radius, $\pm$0.1 mag per sq. arcsec in effective, mean and central surface brightness and $\pm$0.1 in S\'ersic n value. The effective radius (in kpc) assumes a distance of 28.7 Mpc. The fits for IC1459$\_$UDG9 in the r band are more uncertain. 
   \end{table}

   \begin{center}
\begin{table}
 \caption{Coordinates and magnitudes}
\begin{tabular}{lcccc}

\hline
ID & RA         & Dec        & m$_{g}$ & m$_{i}$\\
 & (J2000) & (J2000) & (mag) & (mag)\\
\hline
UDG1  & 22h58m23.09s    &  -36d58m59.0s   &     18.63 & 17.97\\
UDG2  & 22h58m21.25s    &  -36d57m34.9s   &  19.76 & 18.91\\
UDG3  & 22h57m42.36s    &  -36d05m02.0s   &  17.40 & 17.53\\
UDG4  & 22h56m23.27s    &  -36d03m44.0s   &  19.24 & 18.72\\
UDG5  & 22h56m11.95s    &  -37d27m00.0s   &     17.61 & 17.24\\
UDG6  & 22h57m07.04s    & -36d37m31.7s    & 15.83 & 15.38\\
UDG7  & 22h55m49.35s    & -36d38m27.2s    & 16.55 & 16.51\\
UDG8  & 22h56m27.61s    & -36d33m59.5s    & 18.14 & 17.52\\
UDG9  & 22h58m28.74s    & -36d36m02.8s    & 19.71 & 19.68\\
\hline
LSB1  & 22h56m42.09s    & -35d53m55.0s    &    16.15 &  15.50\\
LSB2  & 22h58m30.15s    &  -36d42m08.3s   &  16.58 & 16.13\\
LSB3  & 22h56m14.11s    & -36d28m38.5s    & 17.05 & 16.99\\
\hline
\end{tabular}
\end{table}
\end{center}


  
\section{Conclusions}

Using deep imaging, we identify 9 galaxies in the region of the IC1459 group which meet the criteria for ultra diffuse galaxies (UDGs) once measurement uncertainties are taken into account. This number is consistent with that expected from the UDG abundance--halo mass relation. Two UDGs reveal central nuclei with red colours. A third UDG reveals a double nucleus, with the galaxy centre located between the two nuclei. 
This may be the first reported 
case of a UDG with a double nucleus -- its origin is unclear. 
 We fit all 9 UDGs with a S\'ersic profile and a central component if required. 
 All UDGs generally become bluer in their outskirts suggesting negative metallicity gradients. Their mean colours (g--i = 0.6) and mean S\'ersic n values (0.85) are consistent with those expected for UDGs in low density environments from 
 the simulations of Liao et al. (2019). 
 Several UDGs appear to have an excess of compact objects (e.g. globular clusters) but higher resolution imaging is required to quantify this. We also report the detection and structural parameters of three low surface brightness dwarfs that do not quite satisfy the UDG definition. Future work should include measuring redshifts of each galaxy, in order to confirm their  membership of the IC1459 group, and their velocity dispersion to calculate a dynamical mass. This will give us a better understanding of whether these group UDGs are dark matter dominated or even dark matter free.

\section*{Acknowledgements}

     We thank the full VEGAS survey team for their efforts in acquiring the data used in this work. We thank the referee for their helpful comments. DF thanks the ARC via DP160101608. 
This work is based on visitor mode observations taken at the  ESO  La  Silla  Paranal  Observatory  within  the  VST  Guaranteed  Time  Observations,  Programme  ID 095.B-0779(A). BTD acknowledges supports from a Spanish postdoctoral fellowship `Ayudas 1265 para la atracci\'on del talento investigador. Modalidad 2: j\'ovenes investigadores.' funded by Comunidad de Madrid under grant number 2016-T2/TIC-2039. VS acknowledges financial support from the VST project (PI P. Schipani).












\bsp	
\label{lastpage}
\end{document}